\documentclass[showpacs,epsfig,preprintnumbers,amsmath,amssymb,nofootinbib,20pt]{revtex4-1}
\usepackage{graphicx}
\usepackage{dcolumn}
\usepackage{bm}

\begin{document}
\def\d{{\rm d}}
\def\Epos{E_{\rm pos}}
\def\ap{\approx}
\def\eff{{\rm eft}}
\def\L{{\cal L}}
\newcommand{\vev}[1]{\langle {#1}\rangle}
\newcommand{\CL}   {C.L.}
\newcommand{\dof}  {d.o.f.}
\newcommand{\eVq}  {\text{EA}^2}
\newcommand{\Sol}  {\textsc{sol}}
\newcommand{\SlKm} {\textsc{sol+kam}}
\newcommand{\Atm}  {\textsc{atm}}
\newcommand{\Chooz}{\textsc{chooz}}
\newcommand{\Dms}  {\Delta m^2_\Sol}
\newcommand{\Dma}  {\Delta m^2_\Atm}
\newcommand{\Dcq}  {\Delta\chi^2}
\newcommand{\nbb}{$\beta\beta_{0\nu}$ }
\newcommand {\be}{\begin{equation}}
\newcommand {\ee}{\end{equation}}
\newcommand {\ba}{\begin{eqnarray}}
\newcommand {\ea}{\end{eqnarray}}
\def\VEV#1{\left\langle #1\right\rangle}
\let\vev\VEV
\def\e6{E(6)}
\def\10{SO(10)}
\def\21{SA(2) $\otimes$ U(1) }
\def\321{$\mathrm{SU(3) \otimes SU(2) \otimes U(1)}$ }
\def\lr{SA(2)$_L \otimes$ SA(2)$_R \otimes$ U(1)}
\def\422{SA(4) $\otimes$ SA(2) $\otimes$ SA(2)}
\newcommand{\AHEP}{%
School of physics, Institute for Research in Fundamental Sciences
(IPM)\\P.O.Box 19395-5531, Tehran, Iran\\

  }
\newcommand{\Tehran}{%
School of physics, Institute for Research in Fundamental Sciences (IPM)
\\
P.O.Box 19395-5531, Tehran, Iran}
\def\roughly#1{\mathrel{\raise.3ex\hbox{$#1$\kern-.75em
      \lower1ex\hbox{$\sim$}}}} \def\lsim{\roughly<}
\def\gsim{\roughly>}
\def\ltap{\raisebox{-.4ex}{\rlap{$\sim$}} \raisebox{.4ex}{$<$}}
\def\gtap{\raisebox{-.4ex}{\rlap{$\sim$}} \raisebox{.4ex}{$>$}}
\def\lsim{\raise0.3ex\hbox{$\;<$\kern-0.75em\raise-1.1ex\hbox{$\sim\;$}}}
\def\gsim{\raise0.3ex\hbox{$\;>$\kern-0.75em\raise-1.1ex\hbox{$\sim\;$}}}

\title{A model for lepton flavor violating non-standard neutrino interactions}
\date{\today}

\author{Y. Farzan}\email{yasaman@theory.ipm.ac.ir}
\affiliation{\Tehran}
\affiliation{The Abdus Salam ICTP, Strada Costiera 11, 34151, Trieste, Italy}
\begin{abstract}	
We present a model for Lepton Flavor Violating (LFV) neutral current non-standard interactions of neutrinos with matter fields parameterized by $\epsilon_{\alpha \beta}^f$ with $\alpha \ne \beta$. Here, unlike the previous models, the ratios of the  off-diagonal LFV elements of the effective NSI coupling to the diagonal lepton flavor conserving ones ({\it i.e.,} $(\epsilon_{\alpha \beta}^f)^2/(\epsilon_{\alpha \alpha}^f \epsilon_{\beta \beta}^f)$ ) are arbitrary. The model enjoys rich phenomenology, predicting invisible Higgs decay and new meson decay modes observable in upcoming experiments. The model for $\epsilon_{\mu e}^f$ also predicts a $\mu^-$ to $e^-$ conversion rate on nuclei accessible in the planned experiments.  
\end{abstract}

\date{\today}
\maketitle
\section{Introduction}
Since the establishment of lepton flavor violation  in solar and atmospheric neutrino data, a wide program for the measurement of the parameters of neutrino mass matrix has been started and is vehemently going on. We are now entering neutrino precision era with upcoming experiments being sensitive to the small subdominant effects in the neutrino oscillation. These experiments aim to extract the yet unknown neutrino oscillation parameters, especially, the Dirac CP-violating phase, $\delta_{CP}$. This has also instilled wide interest in the neutral current Non-Standard neutrino Interaction (NSI) with matter fields, $f \in \{ e, {\rm quarks}\}$ parameterized as the following effective potential
\be 2\sqrt{2} G_F \epsilon_{\alpha \beta}^f (\bar{\nu}_\alpha \gamma^\mu\nu_\beta )(\bar{f}\gamma_\mu f), \label{eff} \ee
where $\epsilon_{\alpha \beta}^f$ are dimensionless parameters. In the limit $\epsilon_{\alpha \beta}^f\to 0$, the standard model is recovered. In the presence of NSI,
the propagation of neutrinos in a medium will be affected. There is rich literature studying the effects of NSI on different neutrino oscillation experiments \cite{Dev:2019anc}. It has been shown that if $\epsilon_{\alpha \beta}^f$ are relatively large, NSI will induce degeneracies in the parameter space which should be taken into account when the values 
of neutrino mixing parameters are extracted from observation. In particular, it has been shown that NSI can even mimic the effects of $\delta_{CP}$ \cite{Liao:2016hsa}. Moreover, it has been shown that  neglecting NSI may lead to a wrong determination of the $\theta_{23}$ octant \cite{Agarwalla:2016fkh}.

Of course, if $|\epsilon_{\alpha \beta}^f |\ll 1$, its effects will be negligible. The degeneracies  that we mentioned above appear only for large values of $\epsilon_{\alpha \beta}^f$.  The natural question is that whether we can make a viable $SU(2) \times U(1)$ invariant  model which gives  rise to NSI with such large $\epsilon_{\alpha \beta}^f$ without violating the myriad of bounds that already exist.  
As shown in \cite{Farzan:2015doa,Farzan:2015hkd,Heeck,Tortola,Denton:2018dqq}, invoking a $U'(1)$ gauge symmetry with a light gauge boson, $Z'$ with a mass below $\sim 100$ MeV coupled to both quarks and neutrinos, we can build such models. Building models for lepton flavor conserving NSI ({\it i.e.,} $\epsilon_{\alpha \alpha}^f$) is relatively easy as we can just gauge a linear combination of lepton flavors and the Baryon number. However, obtaining sizable lepton flavor violating (LFV) $\epsilon_{
	\alpha \beta}^f$ with $\alpha \ne \beta$ is more challenging \cite{Amgezi}. If the NSI does not break $SU(2)\times U(1)$, the corresponding charged leptons also receive LFV couplings to $Z'$, leading to fast 
$l^-_\alpha \to l^-_\beta Z'$ at a tree level with a rate enhanced by
$(m_{l_\alpha^-}/m_{Z'})^2$ due to the longitudinal component of $Z'$ \cite{Farzan:2015hkd}. This  problem is overcome in \cite{Heeck} by introducing a new fermion singlet under SM but charged under the new 
$U'(1)$ which mixes with active neutrinos $\nu_\alpha$ and $\nu_\beta$
through a new Higgs doublet whose Vacuum Expectation Value (VEV)
 breaks
 both $SU(2) \times U(1)$ and $U'(1)$. Within this model, LFV NSI can be achieved; however, a relation between the diagonal and off-diagonal
 elements holds
 \be \label{equality} \epsilon_{\alpha \beta}^f\epsilon_{ \beta\alpha}^f=\epsilon_{\alpha \alpha}^f \epsilon_{\beta \beta}^f. \ee
 As discussed in  \cite{Heeck,Tortola}, if we generalize the model to include more than one new fermion mixed with $\nu_\alpha$ and $\nu_\beta$, the Schwartz inequality still implies 
 \be \label{inequality} \left| \epsilon_{\alpha \beta}^f\right|^2<|\epsilon_{\alpha \alpha}^f \epsilon_{\beta \beta}^f|. \ee
 From the phenomenological point of view, the off-diagonal $\epsilon_{\alpha \beta}^f$ elements are  distinguishable from diagonal elements. Ref \cite{Palazzo:2011vg} shows that non-zero $\epsilon_{e\mu}$ and $\epsilon_{e\tau}$ provide a better fit to solar neutrino data than the standard $\nu$ oscillation scheme. Obviously, this  solution does not respect the relation in Eq. (\ref{inequality}).
 
 The aim of the present paper is to build a model in which this inequality is violated: $|\epsilon_{\alpha \beta}^f| > |{\epsilon_{\alpha \alpha}^f \epsilon_{\beta \beta}^f}|^{1/2}.$ 
 In sec \ref{model}, we present the underlying model for the LFV NSI and discuss the bounds that already  exist on the parameters of the model. In sec. \ref{outlook},
  we summarize our results and suggest strategies to test the model.
\section{The model \label{model}}
In this section, we build an underlying model for effective Lagrangian in Eq. (\ref{eff}). Our model contains a new $U(1)$ gauge boson $Z'_\mu$ which couples both to the  matter fields $f$ ($f=e, u$ or $d$) and to neutrinos as follows
\be g_f (\bar{f}\gamma^\mu f)Z'_\mu \label{fZ} \ee and
\be (g_\nu)_{\alpha \beta} (\bar{\nu}_\alpha \gamma^\mu\nu_\beta ) Z'_\mu  \ \ \ \ \ \alpha \ne \beta . \label{nuZ}
\ee
For the low energy-momentum transfer, we can then write
\be \label{relation} \epsilon^f_{\alpha \beta}=\frac{g_f (g_\nu)_{\alpha \beta}}{2 \sqrt{2} G_F m_{Z'}^2} . \ee
 In sect. \ref{off-sec}, we introduce the underlying scenario that leads to the off-diagonal coupling of  (\ref{nuZ}) in two versions: First in minimal version that violates the lepton number and then in the lepton number conserving case. We then discuss the bounds on the parameters of this sector of the model and briefly comment on the prospects for testing the model in the future.
 In sect \ref{kin-sec}, we discuss the model for the interaction of Eq. (\ref{fZ}) and the phenomenological  consequences of this model.
 
\subsection{ An electroweak invariant model for $Z'_\mu \bar \nu_\alpha \gamma^\mu \nu_\beta$ with $\alpha \ne \beta$ and without $Z'_\mu \bar l_\alpha \gamma^\mu l_\beta$ \label{off-sec}}
In sec. \ref{lnv}, we introduce the minimal version of the model which breaks lepton number and may therefore induce large contribution to neutrinoless double beta decay and/or lepton number violating processes such as $B^- \to \pi^+ \mu^- \mu^-$. In sec. \ref{lnc}, we show that with a slight change in the content of the model, a lepton number conserving model can be built without large contribution to neutrinoless double beta decay and other lepton number violating processes. In sec. \ref{bounds}, we shall review the experimental bounds on the parameters of model. We will discuss which part of the parameter space gives us large enough $g_\nu$.
  
\subsubsection{Lepton number violating version of the model \label{lnv}}
Let us first introduce two right-handed Weyl fermions which are singlets of the standard model
gauge  group but 
under the new gauge  $U'(1)$ transform as   $\psi_1 \to e^{i \alpha}\psi_1$ and $\psi_2 \to e^{-i \alpha}\psi_2$. (In sec \ref{lnc}, we shall discuss another version of the model in which $\psi_1$ and $\psi_2$ are promoted to be Dirac fermions.) Having opposite charges, the $U'(1)$ gauge anomaly will be canceled. Moreover, we can write mass term for them as 
\be \frac{M_N}{2} (\psi_1^T c\psi_2+\psi_2^T c\psi_1)+H.c.=\frac{M_N}{2}
(N_1^TcN_1-N_2^TcN_2)+H.c. \label{LNVmass}\ee
where in the right hand side of the equation, we have replaced
\be \psi_1\equiv \frac{N_1+N_2}{\sqrt{2}} ~~~~~ {\rm and}~~~~~  \psi_2\equiv \frac{N_1-N_2}{\sqrt{2}} \label{decompose}, \ee
so $N_1$ and $N_2$ have equal masses.
The gauge interaction can be then written as
\be g_\psi (\bar{\psi_1}\gamma^\mu \psi_1-\bar{\psi_2}\gamma^\mu \psi_2)Z'_\mu=g_\psi (\bar{N}_1 \gamma^\mu N_2+\bar{N}_2 \gamma^\mu N_1)Z'_\mu.
\ee

Notice that $Z'$ converts $N_1$ and $N_2$ to each other.
If we construct a mechanism that mixes $N_1$ only with $\nu_\alpha$ and $N_2$ only with $\nu_\beta$, we obtain an interaction of type in Eq. (\ref{nuZ}) without diagonal couplings of form $\bar \nu_\alpha \gamma^\mu \nu_\alpha Z'_\mu$ or $\bar \nu_\beta \gamma^\mu \nu_\beta Z'_\mu$ and without corresponding LFV couplings for charged leptons, $\bar l_\beta \gamma^\mu l_\alpha Z'_\mu$. In the following, we show that it is possible to build a model for such mixing after spontaneous gauge symmetry breaking.
To do so, we shall invoke the famous inverse seesaw mechanism \cite{inverse} for neutrino mass generation involving $\tilde{N}_{1L}$, $\tilde{N}_{2L}$, $\tilde{N}_{1R}$ and $\tilde{N}_{2R}$ where $(\tilde{N}_{1L},\tilde{N}_{1R})$ and $(\tilde{N}_{2L},\tilde{N}_{2R})$ form Dirac fermions.
In our model, these Weyl fermions are all singlets both under the standard
 model gauge group and the new $U'(1)$ gauge group. In our model, the $U'(1)$ is broken by the VEV of a new scalar $\phi$ which is a singlet of the SM gauge group and its $U'(1)$ charge is equal to that of $\psi_1$ and opposite to that of $\psi_2$; {\it i.e.,} under $U'(1)$, $\phi \to e^{i \alpha} \phi$.

Let us introduce a $Z_2$ symmetry under which, $$\psi_1 \leftrightarrow \psi_2,  \  \ \ \ \ \ \  \ \phi  \leftrightarrow\phi^*  \ \ \ \ \ \ {\rm and} \ \ \ \ \ \ Z' \to -Z'.$$ The  parities of $\tilde{N}_i$, $L_\alpha$ and $L_\beta$ are shown in table \ref{tab}. The other standard model particles are parity even under this $Z_2$.
\begin{table}[htb]
	\begin{center}
		\begin{tabular}{|c|c|c|c|c|c|}
 \hline 
 $\tilde{N}_{1 L}$ & $\tilde{N}_{1 R}$ &$ L_\alpha$ &  $\tilde{N}_{2L}$ & $\tilde{N}_{2 R}$ &$ L_\beta$ \cr  \hline
 1&1& 1 &-1 &-1&-1\cr
 \hline
 \end{tabular}
\caption{ $Z_2$ parities.\label{tab}}
\end{center}
\end{table}

As we mentioned above, our model invokes inverse seesaw mechanism with the following potential which respects $Z_2$
\be \lambda_\alpha \bar{\tilde{N}}_{1R} H^TcL_\alpha +M_1 \bar{\tilde{N}}_{1L}\tilde{N}_{1R}+\lambda_\beta \bar{\tilde{N}}_{2R} H^TcL_\beta +M_2 \bar{\tilde{N}}_{2L}\tilde{N}_{2R} +{\rm H.c.}\label{LC}\ee Let us define
$$\tilde{m}_\beta =\lambda_\beta v/\sqrt{2} \ \ \ \ {\rm and} \ \ \ \  \tilde{m}_\alpha =\lambda_\alpha v/\sqrt{2}.$$ Assigning lepton number equal to 1 to $\tilde{N}_i$, it is easy to show that the potential in Eq. (\ref{LC}) is    lepton number conserving and does not induce a mass for light neutrinos. To obtain mass for active neutrinos, the following lepton number violating masses should be added: 
\be \mu_{1L }\tilde{N}_{1L}^T c\tilde{N}_{1L}+\mu_{2L}
\tilde{N}_{2L}^T c\tilde{N}_{2L}+(L\leftrightarrow R)+H.c. \label{main-mu} \ee plus
\be \mu_{21L }\tilde{N}_{2L}^T c\tilde{N}_{1L}+(L\leftrightarrow R). \label{z2-mu} \ee
Including these terms, SM neutrinos will obtain a mass proportional to  the $\mu$ terms. This is the basis of the so-called inverse seesaw mechanism. Notice that $\mu_{21}$ in Eq (\ref{z2-mu}) breaks  $Z_2$ making it possible to obtain a mixing between the $\alpha$ and $\beta$ flavors in the neutrino mass matrix. We shall not elaborate further on neutrino masses as the inverse seesaw mechanism is widely studied in the literature.

To obtain a mixing between $N_i$ and the neutral leptons, we assign lepton number of $-1$ to $N_1$ and $N_2$ (or equivalently to $\psi_1\equiv (N_1+N_2)/\sqrt{2}$ and to $\psi_2\equiv (N_1-N_2)/\sqrt{2}$)  and introduce the following $Z_2$ and gauge invariant coupling:
\be \frac{Y_1}{\sqrt{2}}(\phi \bar\psi_1+\phi^* \bar\psi_2)c\tilde{N}_{1R}^*+
	\frac{Y_2}{\sqrt{2}}( \phi \bar\psi_1-\phi^* \bar\psi_2)c\tilde{N}_{2R}^*+{\rm H.c.} 
		\ee
	The vacuum expectation value of $\phi$,	$v_\phi=\langle \phi \rangle$, breaks the $U'(1)$ symmetry and therefore induces mass  mixing terms  as 
		$$m_1  N_1^\dagger c\tilde{N}_{1R}^*+m_2  N_2^\dagger c\tilde{N}_{2R}^*+{\rm H.c.}$$
		where $m_1= Y_1 v_\phi$ and $m_2= Y_2 v_\phi$.
		
		Notice that the $M_N$ mass term in Eq. (\ref{LNVmass}) is an explicit source of lepton number violation. Despite this source of lepton number violation, 
		in the limit  that the $\mu$ terms in Eqs. (\ref{main-mu}) and (\ref{z2-mu}) vanish,
		SM neutrinos will remain massless. This can be understood because, in the limit of $\mu \to 0$, the symmetric mass matrix for ($\nu_\beta, \tilde{N}_{2L}, c\tilde{N}_{2R}^*, c N_2^*$) can be written as
		\begin{eqnarray} \left[ \begin{matrix} 0 & 0 & \tilde{m}_\beta & 0 \cr 0 & 0  & M_2 & 0 \cr \tilde{m}_\beta & M_2 & 0 & m_2 \cr 0 & 0 & m_2
		& M_N \end{matrix} \right] \end{eqnarray} whose determinant vanishes so the lightest mass eigenvalue  will vanish independent of the values of $m_2$, $M_2$ or $M_N$. 
			 The light neutrino will be a linear combination as $$\nu_\beta+\sin \beta'\tilde{N}_{2L}  +\sin \beta N_2, $$ where $\beta$ and $\beta'$ are small mixing angles. 
			  Similar consideration holds valid for the mass matrix of $\nu_\alpha$, $\tilde{N}_1$ and $N_1$. We can similarly write $$ \nu_{\alpha}+\tilde{N}_{1L }\sin \alpha'+N_1 \sin \alpha.$$ We then obtain
			  \be (g_\nu)_{\alpha \beta} =g_\psi \sin \alpha \sin \beta.\ee
			  We demand the masses of $\tilde{N}_i$ and $N_i$ to be larger than $\sim 500$ MeV to avoid the bounds from supernova type II cooling and meson decay. This can be achieved if $M_2,M_1,
			  M_N> 500$ MeV. The bounds from low energy experiments (such as meson decays) set an upper bound on $\sin^2\beta +\sin^2 \beta'$ and on  $\sin^2\alpha +\sin^2 \alpha'$. In the limit $M_1\gg m_1$ ($M_2\gg m_2$), we find $\sin \alpha' \gg \sin \alpha$ ($\sin \beta' \gg \sin \beta$). In order for $\sin\alpha$ and $\sin \beta$ to saturate the bounds and therefore to obtain largest possible $g_\nu$, we focus on the range that $m_2 \sim M_2$ and $m_1 \sim M_1$. Remembering that $m_1$ and $m_2$ 
			  are given by  $\langle \phi\rangle $ which also contribute to the $Z'$ mass, we find
			  \be m_{Z'}\sim M_{1,2} g_\psi/Y_{1,2}>  500~ {\rm MeV} g_\psi \ . \label{Z'bound} \ee

			  The following remarks are in order.
			  
			  \begin{itemize} \item Notice that the $Z_2$ symmetry guarantees that $N_i$ mixes with only one of $\nu_\alpha$ or
			  $\nu_\beta$ and as a result while we obtain LFV coupling $Z'_\mu \bar{\nu}_\alpha \gamma^\mu \nu_\beta$, we do not obtain LFC couplings of 
			  $Z'_\mu \bar{\nu}_\alpha \gamma^\mu \nu_\alpha$ and  
			  $Z'_\mu \bar{\nu}_\beta \gamma^\mu \nu_\beta$. As a result, the bound $|\epsilon_{\alpha \beta}|^2 \leq |\epsilon_{\alpha \alpha} \epsilon_{\beta \beta}|$ within the model(s) in Ref \cite{Heeck} does not apply here.
			  Without the $Z_2$ symmetry, $N_1$ and $N_2$ could mix simultaneously with $\nu_\alpha$ and $\nu_\beta$ leading to   $Z'_\mu \bar{\nu}_\alpha \gamma^\mu \nu_\alpha$ and  
			  $Z'_\mu \bar{\nu}_\beta \gamma^\mu \nu_\beta$ along with $Z'_\mu \bar{\nu}_\alpha \gamma^\mu \nu_\beta$ .
			  \item We could obtain a mixing between active neutrinos and $N_i$ in a more economic version of the model without introducing $\tilde{N}_{iL}$ (with a lepton number violating mass of the form $\tilde{N}_{iR}^T c\tilde{N}_{i R}$) but in this case the active neutrinos would obtain a mass of order $m_{N} \sin^2 \alpha \sim m_N \sin^2 \beta$ which for $m_N \sim 500$ MeV would imply $\sin \alpha, \sin \beta <10^{-5}$ rendering $\epsilon_{\alpha \beta}$ too small.
			  \end{itemize}
			  
			  \subsubsection{Lepton number conserving version of the model \label{lnc}}
			 Promoting $\psi_1$ and $\psi_2$ to Dirac fermions, we can impose lepton number conservation up to small effects induced by the $\mu$ terms in Eqs. (\ref{main-mu}) and (\ref{z2-mu}). The $Z_2$ symmetry then implies the mass terms for $\psi_1$ and $\psi_2$ to be of form
			 \be m_N(\bar{\psi}_{1L}{\psi}_{1R}+\bar{\psi}_{2L}{\psi}_{2R}+{\rm H.c.})=
			  m_N(\bar{N}_{1L}{N}_{1R}+\bar{N}_{2L}{N}_{2R}+{\rm H.c.}) .\ee
			  The rest of features of the model will be similar to the model described in sec \ref{lnv}.
			  	\subsubsection{ Bounds on the model parameters\label{bounds}}
			  	In this section, we  discuss the bounds on the mixing of sterile neutrinos with active neutrinos and other parameters of the model. We also comment on the possibility to test the model. For sterile neutrinos lighter than a few MeV, there are strong bounds on the mixing from cosmology \cite{Dolgov:2000jw}.
			  	For masses below $\sim 100$~MeV, there are strong constraints from the supernova cooling. For masses below that of Kaon ($\simeq 500$ MeV), strong bounds on the mixing come from the Kaon and pion decay. We therefore assume the masses of sterile neutrinos, determined by $m_N$ and $M_i$, to be heavier than 500~MeV. 
			  	
			  	In the literature, strong bounds on the 
			  	mixing of sterile neutrinos with mass heavier than 500 MeV with active neutrinos have also been reported from NuTeV \cite{Vaitaitis:1999wq}, WA66 \cite{CooperSarkar:1985nh}, CHARM II \cite{Vilain:1994vg}, BELLE \cite{Liventsev:2013zz}, Higgs decay \cite{Higgs}, NA62 \cite{NA62}, L3 \cite{L3}, DELPHI \cite{delphi}  and ATLAS+CMS
			  	\cite{Aad:2019kiz}.
			  	Ref. \cite{Dev} gives an updated compilation of the bounds and a forecast for future searches (see also, \cite{Silvia}).
			  A sterile neutrino mixed with $\nu_\alpha$ can decay into $\nu_\alpha \nu_\gamma \bar{\nu}_\gamma$ and, if the kinematics allows, into $l_\alpha \bar l_\gamma \nu_\gamma$ (where $\gamma$ may or may not correspond to $\alpha$) via electroweak interaction with a rate suppressed by the square of the mixing. The bounds that we enumerated are all based on searches for the signature of the final charged particles. In our model, there is a possibility of faster two body decay into  active neutrinos and $Z'$. If $Z'$ decays into a neutrino pair, it will not show up in these experiments so all these bounds can be avoided. 
			  Let us formulate the condition for avoiding the bounds. Notice that in our model, in addition to $N_1$ and $N_2$, we have $\tilde{N}_1$ and $\tilde{N}_2$ which also mix with the active neutrinos. In order to open the decay mode into $Z'$ for $\tilde{N}_1$ and $\tilde{N}_2$ (or to be more precise for mass eigenstates composed mainly of $\tilde{N}_i$ and $N_i$), we need a large mixing between them. In sect \ref{lnv}, we already showed that $m_i \sim M_i$. If we further impose the condition $m_i \sim M_i \sim M_N$, this condition will be fulfilled and the decay rates of all these sterile neutrinos into $Z'$ and active neutrinos will be of the same order.
			  In order for the two body decay into $Z' \nu_a$ to dominate over the electroweak three body decay, we just need  $g_\psi^2 \gg (g_{SU(2)}^4/16 \pi^2) (m_N/m_W)^4$ which can be readily satisfied. A more challenging requirement is that $Z'$ dominantly decays into neutrinos.  This requires 
			  $g_\nu^2 \gg g_e^2$, for $m_{Z'}> 2 m_\pi$  $g_\nu^2\gg g_q^2$, 
			  for $m_{Z'}> 2 m_\mu$  $g_\nu^2\gg g_\mu^2$ and for $m_{Z'}> 2 m_\tau$  $g_\nu^2 \gg g_\tau^2$. In summary, in order to avoid the enumerated bounds on the mixing of sterile neutrinos with mass above 500 MeV, the decay mode of the sterile neutrinos into $Z' \nu_a$ and then $Z' \to \nu_a \bar{\nu}_a$ must dominate. This in turn requires $m_N \sim m_i \sim M_i$ and $g_\nu^2 \gg g_f^2$.

			  	In case that the mass of sterile neutrino is of Majorana type, the null results from
			  	neutrinoless double beta decay searches  set a strong  bound on the mixing with $\nu_e$, ranging from few$\times 10^{-5}$ to  few$\times 10^{-2}$ for $500~{\rm MeV}<M_N  \lsim 10 ~{\rm TeV}$. If we are interested in $\epsilon_{e \tau}$ or $\epsilon_{e \mu}$ close to the present bounds, we then need to adopt the lepton number conserving version of the model as described
			  	in sec. \ref{lnc} to avoid the bounds from $0\nu \beta \beta$. 
			  	Moreover, for the Majorana type sterile neutrinos, there is also a strong bound on the mixing with $\nu_\mu$ of order of $10^{-2}$ from searches for lepton number violating decay mode $B^-\to \pi^+ \mu^-\mu^-$ from the LHCb
			  	\cite{Aaij:2014aba}.
			  	
			  	Since active neutrinos mix with sterile neutrinos, the PMNS $3 \times 3$ matrix will not be unitary. There are strong bounds on the 
			  	 violation of the unitarity of the PMNS matrix \cite{unitarity-violation}. Some of these bounds do not however apply to our case. Most notably despite the deviation of the PMNS matrix from unitarity, in our model we do not obtain a significant contribution to $l_\alpha^- \to l_\beta^- \gamma$ at one loop level as $\nu_\alpha$ and $\nu_\beta$ mix with different sets of sterile neutrinos. In other words, the heavier mass eigenstate either have a contribution from $\nu_\alpha$ or from $\nu_\beta$ but not from both, making one loop contribution absent. As a result, the bounds on 
			  	$(U_{PMNS}^\dagger \cdot U_{PMNS})_{\alpha \beta}|_{\alpha \ne \beta}$ 
			  	from $l_\alpha \to l_\beta \gamma$ discussed in the literature \cite{unitarity-violation} does not apply here. There will be a two-loop contribution in which both $W$ and $Z'$ propagate but the effect will be both GIM and two-loop suppressed and therefore negligible. 
			  	
			  	 As long as 
			  	 the heaviest sterile neutrino mixed with the active neutrinos is much lighter than $m_Z/2$, the bounds on the invisible decay width of $Z$ and those from the leptonic decay modes of $W$ can also be relaxed in our model. In other words, the deviation from standard model prediction for ($Z\to$invisibles) and for ($W \to l$+missing energy) will be suppressed not only by the square of mixing but also by $O(m_N^2/ m_W^2)$. However for $m_N>m_K$, $K^+ (\pi^+) \to l_\alpha^+ \nu$ and
			  	  $K^+ (\pi^+) \to l_\beta^+ \nu$ will be suppressed by $\cos^2 \alpha$ and $\cos^2 \beta$, respectively. Moreover, the rate of the muon decay which is used to extract $G_F$ will be affected. To be on the safe side, we take 
			  	  $$ \sin \alpha \sin \beta <10^{-3}$$
			  	  to satisfy the bounds from the violation of the unitarity of the PMNS matrix \cite{unitarity-violation}.
			  	   We then find
			  	  \be (g_\nu)_{\alpha \beta} |_{\alpha \ne \beta}<10^{-3} g_\psi \ . \label{GNU}\ee
			  	  There are also direct bounds on $(\sum_{\alpha}|(g_\nu)_{e\alpha}|^2)^{1/2}$ and
			  	  $(\sum_{\alpha}|(g_\nu)_{\mu\alpha}|^2)^{1/2}$ from
			  	  the $K^+$ and $\pi^+$ decays into $e^+$ and $\mu^+$ plus missing energy which are again around $10^{-3}$ \cite{Pouya1}. As shown in \cite{Pouya2}, the DUNE near detector  can probe small values of $g_\nu$ and even determine its flavor structure.
			  	  
			  	  As discussed before, the contribution to $l_\alpha \to l_\beta \gamma$ in our model is two-loop suppressed but at one loop, we obtain 
			  	  $l_\alpha \to l_\beta Z'$ with a rate estimated as 
			  	  $$ \frac{m_\alpha}{4 \pi}\left( \frac{m_\psi^2}{m_W^2}\frac{g_\nu}{16 \pi^2}\right)^2 g_{SU(2)}^4 \frac{m_\alpha^2}{m_{Z'}^2}\ , $$
			  	  where $m_\psi^2/m_W^2$ comes from the GIM suppression and $(m_\alpha^2/m_{Z'}^2)$ is the enhancement factor due to the longitudinal component of $Z'$. In PDG \cite{pdg}, there are explicit bounds on such exotic decay modes of $\tau$:
			  	  $$ {\rm Br}(\tau \to e Z')<2.7\times 10^{-3}  \ \ \ {\rm and}  \ \ \ 
			  	   {\rm Br}(\tau \to \mu Z')<5\times 10^{-3} \ . $$
			  	   These bounds can be easily satisfied for $g_\nu <10^{-3}$ and $m_{Z'}>10$~MeV. 
			  	  The present bound on Br($\mu \to e Z'$) is of order of $10^{-5}$ \cite{heeck}. Taking $(g_\nu)_{\mu e}<10^{-3}$, $m_\psi \sim$ GeV and $m_{Z'}\sim 10$ MeV, we find that  ${\rm Br}(\mu \to e Z')$ is of similar order.  Future muon decay experiments can easily test this prediction \cite{heeck}.
			  	   
			  	   Similarly, at one loop level and through $Z'$ exchange, $
			  	 \mu$ to $e$ conversion can take place with
			  	 \be R = \frac{\Gamma (\mu +N \to e +N)}{\Gamma (\mu +N \to \nu_\mu +N')}
\sim \frac{(g_\nu)_{e \mu}^2 g_q^2}{(16 \pi^2)^2}\left( \frac{m_\psi^2}{m_\mu^2+m_{Z'}^2}\right)^2=5\times 10^{-15} 	\left(\frac{g_\nu}{10^{-3}}\right)^2 \left(\frac{g_q}{10^{-4}}\right)^2	\left( \frac{m_\psi}{1~{\rm GeV}}\right)^4 . \ee	  	   
The present bound is set by SINDRUM II collaboration which is $7 \times 10^{-13}$ \cite{sindrum} so the present bound can be easily satisfied. The next generation Mu2e and
COMET experiments can probe $R$ down to $5 \times 10^{-17}$ \cite{Mu2e} so they can be sensitive down to $(g_\nu)_{e \mu} g_q \sim 10^{-8}$.

In our model, the Higgs will have new invisible decay modes into $\bar{\nu}_\beta \tilde{N}_2$ and 
$\bar{\nu}_\alpha \tilde{N}_1$ with  rates of $\lambda_\beta^2 m_H /(4\pi)$ and $\lambda_\alpha^2 m_H /(4\pi)$, respectively. Taking $\lambda_{\alpha,\beta} \langle H \rangle=\tilde{m}_{\alpha},\tilde{m}_{ \beta}<2$ GeV, the present bound on the branching ratio of the invisibles decay mode, Br($H \to {\rm invisibles})<0.2$ \cite{pdg}, can be satisfied. Future searches for $H \to {\rm invisibles}$ can test the model. In fact, if Br($H \to {\rm invisibles}$) down to $O(1 \%)$ is measured, the entire parameter space of interest to us can be probed. In principle, we can consider the masses of the sterile neutrinos to be heavier than the Higgs mass but then the mixing parameters $\sin \alpha$ and $\sin \beta$ will be suppressed by $m_{1,2}/m_N\sim (m_{Z'}/m_N)(Y_{1,2}/g_\psi)$. (Remember that we require relatively light $Z'$ to obtain sizable NSI.) 
			 \subsection{ Couplings of matter fields to $Z'$\label{kin-sec}}

In this section, we discuss two mechanisms for coupling $Z'$ to matter fields and then discuss the bounds on the coupling. We also evaluate the maximum values of $\epsilon_{\alpha \beta}^f$ that can be achieved, combining these bounds with the bounds on $g_\nu$ discussed in sec \ref{bounds}. We then propose ideas to test the model.
\begin{itemize} \item
	Like Refs. \cite{Farzan:2015doa,Farzan:2015hkd,Heeck,Tortola,Denton:2018dqq}, in order to couple $Z'$ to matter fields, we may identify		 
the $U'(1)$ gauge symmetry with \be B-(a_e L_e +a_\mu L_\mu +a_\tau L_\tau) \ee  with a coupling of $g_B$. Thus, $g_q=g_B/3$ and $g_\alpha =a_\alpha g_B$ where $\alpha \in \{e, \mu , \tau \}$. Notice that such gauge symmetry will 
also lead to lepton flavor conserving coupling $ g_B a_\gamma Z_\mu' \bar{\nu}_\gamma  \gamma^\mu \nu_\gamma $. 
Taking $a_\gamma g_B\ll g_\nu$, $\epsilon_{\alpha \beta}$ will be larger than $\epsilon_{\gamma \gamma}$. For $a_e+a_\mu+a_\tau=3$, the gauge anomalies cancel out and there is no need to add new chiral doublets to cancel the $U'(1)-SU(2)-SU(2)$ anomaly. However, if $a_e+a_\mu+a_\tau\ne 3$, heavier doublets need to be added to cancel the anomalies.
For further details and discussions, see  \cite{Farzan:2015doa,Farzan:2015hkd,Heeck,Tortola,Denton:2018dqq}.
\item
Another possibility is  a kinetic mixing between $Z'$ and the hypercharge gauge $B_\mu=\cos \theta_W A_\mu -\sin \theta_W Z_\mu$ as \be \delta Z_{\mu \nu}' B^{\mu \nu} \ee where $Z_{\mu \nu}'$ and $B_{\mu \nu}$ are field strength of these gauge bosons.  Going to the canonical basis where both gauge boson mass terms and the kinetic terms are diagonal and properly normalized, we find that $Z'$ obtains a vector-like coupling to charged fermions which in the leading order in $\delta$ is given by 
\be g_f=(Q_f e)	\cos \theta_W \delta . \ee As expected, the axial part of the coupling to charged fermions as well as the coupling to neutrinos, which originate from the contribution of $Z_\mu$ to
$B_\mu$ (rather than that of $A_\mu$ to $B_\mu$), are further suppressed by a factor of $ (m_{Z'}^2/m_Z^2)\delta \sin \theta_W$. 
 Remembering that the Deuteron dissociation process that is invoked by SNO to measure the rate of the neutral current interaction of solar neutrinos is only sensitive to the axial current, no significant bound from this measurement applies to our model.
 Up to $O(\delta)$, $N_1$ and $N_2$ will not obtain couplings to the photon so we do not need to worry about the bounds on the millicharged particles.\footnote{As shown in \cite{Liu}, to obtain milli-charged particles, in addition to the kinetic mixing, a mass mixing term between $B_\mu$ and $Z'_\mu$ is required.
 	} Notice that within this scenario, the effective coupling, $\epsilon^f$ will be proportional to the electric charge of the matter field, $Q_f$. This means that for neutrinos propagating in neutral media ({\it i.e.,} all the media such as Earth, Sun, supernova and etc that we know of), there will be no NSI effect on the propagation. However, for the coherent elastic neutrino nucleus scattering experiments such as COHERENT \cite{Akimov:2017ade} or CONUS \cite{Farzan:2018gtr}, the effects of NSI will be present and proportional to the square of the atomic number. The appearance of NSI effects in these experiments and the lack of evidence for them in the
 	 neutrino propagation experiments such as DUNE and atmospheric or solar neutrino experiments can be indicative of this particular scenario in which the coupling of the mediator to the matter field originates from the kinetic mixing.
 \end{itemize}
 
 Let us now discuss the experimental bounds. For $Z'$ lighter than $\sim 100$ MeV, there are strong bounds on $g_e$ from beam dump experiments \cite{NA64} and supernova cooling. As a result, for the case that $a_e \ne 0$ or in the kinetic mixing scenario, we should take
 $Z'$ to be heavier than $\sim 100$ MeV. Still we will have strong bounds from neutrino electron scattering experiments  such as BOREXINO, GEMMA and CHARM but these bounds are already encoded in the bounds on NSI parameter, $\epsilon^e_{\alpha \beta}$. 
 As discussed before, the condition $g_e \ll g_\nu$ has to be fulfilled to guarantee that the invisible decay mode of $Z'$ dominates. Taking $(g_\nu)_{\alpha \beta}=10^{-3} g_\psi$, $m_{Z'}=500~{\rm MeV} g_\psi$ (see Eq. \ref{Z'bound}) and $g_e =0.1 g_\nu$, we find $\epsilon_{\alpha \beta}^e=0.01 ((g_e/g_\nu)/0.1)$ which is close to the upper bounds on  $\epsilon_{\alpha \beta}^e$ \cite{Tortola}. In the kinetic model, $\epsilon_{\alpha \beta}^p=-\epsilon_{\alpha \beta}^e$ and in the Baryon
number gauging model, $\epsilon_{\alpha \beta}^q=-\epsilon_{\alpha \beta}^e/(3 a_e)$
which again is close to the present upper bounds \cite{Tortola,Denton:2018xmq,Demidov:2019okm,Coloma:2019mbs,Arman}.

 In the scenario where we gauge a linear combination of lepton and Baryon numbers, if we set $a_e =0$, a wider mass range for $Z'$ will open. $Z'$ lighter than $\sim 10 $ MeV is disfavored by cosmology \cite{Huang:2017egl}. For  $130~{\rm MeV}>m_{Z'}>10~{\rm MeV}$ we can have $g_q \sim g_\tau\sim g_\mu\sim g_\nu <10^{-3}$ without violating any bound \cite{Harnik:2012ni}. Notice that for this mass range even if $g_q,g_\mu,g_\tau>g_\nu$,   $Z' \to \nu \bar{\nu}$ will be still the dominant decay mode. 
Taking $g_q =10^{-3}$ (which saturates the bound from $\pi^0 \to \gamma Z'$
\cite{Gninenko:1998pm}), $(g_\nu)_{\alpha \beta}=10^{-3} g_\psi$ and $m_{Z'}\sim 500~{\rm MeV}g_\psi$ (with $g_\psi>0.02$ to make $m_{Z'}$ heavier than 10 MeV and therefore avoid the cosmological bounds), we obtain $$
\epsilon^q_{\alpha \beta}=\frac{0.12}{g_\psi}\frac{g_q}{10^{-3}}
$$  so the values of $\epsilon_{\alpha \beta}^q$ can easily saturate the present bounds on them. In this model, we also obtain diagonal NSI couplings as
\be \epsilon_{\mu \mu}^q=-3 \frac{a_\mu g_q}{(g_\nu)_{\alpha \beta}} \epsilon_{\alpha \beta}^q \ \ \ \ \ {\rm and } \ \ \ \ \ \epsilon_{\tau \tau}^q=-3 \frac{a_\tau g_q}{(g_\nu)_{\alpha \beta}} \epsilon_{\alpha \beta}^q .\ee
To have anomaly cancellation we should have $a_\mu+a_\tau=3$ so $a_\mu$ and $a_\tau$ cannot be simultaneously zero unless new chiral doublets are added to the model to cancel the anomalies. We can however have $a_\mu=0$ and $a_\tau=3$ (or $a_\tau=0$ and $a_\mu=3$). In this case, it is obvious that for any $\alpha \ne \beta$, $|\epsilon_{\alpha \beta}^q|^2 >|
\epsilon_{\alpha \alpha}^q \epsilon_{\beta \beta}^q|$. More interestingly, since we have a freedom in the choice of $a_\mu/a_\tau$ and $g_q/(g_\nu)_{\alpha \beta}$, we can obtain an arbitrary ratio of $\epsilon_{\alpha \beta}^q/\epsilon_{\alpha \alpha}^q$. For example, we can have  $\epsilon_{e e}^q=\epsilon_{\mu \mu}^q=0$, a value of  $\epsilon_{e \mu}^q$  saturating the bound and $\epsilon_{\tau \tau}^q\ne 0$ with an arbitrary value of 
$\epsilon_{e \mu}^q/\epsilon_{\tau \tau}^q$

 For heavier $Z'$, visible decay modes $Z' \to \mu \bar{\mu}$ and $Z' \to \pi^+\pi^-$ open so as we discussed in sec \ref{bounds}, we should impose $g_f \ll g_\nu$.
 Like the case of nonzero $g_e$, we can still obtain large $\epsilon_{\alpha \beta}^q$ 
 with $\epsilon_{\alpha \beta}^q \gg \epsilon_{\alpha \alpha}^q ,\epsilon_{\beta \beta}^q$.   
 
 Notice that  in our model, all the new particles, including the new scalar $\phi$, have a mass above $\sim 10$ MeV and decay away well before the neutrino decoupling in the early universe so the model is not constrained by the bound on the extra relativistic degrees of freedom from cosmological constraints.
 \section{Summary and outlook \label{outlook}}
 We have built a $SU(2)\times U(1)$ invariant UV complete model which leads to NSI with LFV couplings $\epsilon_{\alpha \beta}^f$  with $\alpha \ne \beta$ and arbitrary ratios of $\epsilon_{\alpha \beta}^f/\epsilon_{\alpha \alpha}^f$ and $\epsilon_{\alpha \beta}^f/\epsilon_{\beta
 	 \beta}^f$. 
 	The effects of such NSI can show up in various neutrino oscillation experiments and in different coherent elastic neutrino nucleus scattering experiments. The model is based on a pair of fermions which are singlets of the standard model but have opposite charges under a new $U'(1)$ gauge symmetry with a relatively light gauge boson of mass $10~{\rm MeV}<m_{Z'}< {\rm few}~100~{\rm MeV}$. The lower bounds on $m_{Z'}$ is imposed by BBN and CMB \cite{Huang:2017egl}. After electroweak and $U'(1)$ symmetry breaking, the new fermions mix with the ordinary neutrinos of different flavors leading to LFV couplings that we require. The model incorporates the inverse seesaw mechanism which helps to decouple the scales of the masses of the new sterile neutrinos from the masses of active neutrinos. The values of $\epsilon_{\alpha \beta}^f$ are given by the mixing between the sterile neutrinos and the active ones.  We have presented the model in two versions: lepton  number violating version and lepton number conserving one. We have also proposed two different mechanisms to couple $Z'$ to the matter fields based on gauging a linear combination of the Baryon and lepton numbers and based on a kinetic mixing between the field strengths of the new gauge boson and that of the hypercharge. We have discussed how we can distinguish them by combining the results from the  neutrino oscillation experiments and from the coherent elastic neutrino nucleus scattering experiments.
 
 We have taken the masses of the sterile neutrinos to be heavier than 500~MeV ({\it i.e.,} $m_{K^+}$) to avoid the strong bounds on their mixing from the Kaon and pion decay. In our model, the sterile neutrinos promptly decay into $Z'$ and SM neutrinos and appear as missing energy in colliders so the bounds that exist on the mixing of the sterile neutrinos with mass above 0.5 GeV from searching for their charged decay products do not apply here.
 On the other hand, we discuss that since the active sterile mixing, $\sin \alpha$ and $\sin\beta$ are expected to be of order of the ratio of $m_{Z'}$ to the masses of sterile neutrinos, large $\epsilon_{\alpha \beta}$ (being proportional to $\sin \alpha \sin \beta /m_{Z'}^2$) implies that the sterile neutrino masses are not  much heavier than a few GeV.  The invisible decay of the Higgs also imposes a similar bound. The masses of sterile neutrinos should be in the range between 500 MeV to a few GeV. The entire range can be probed by improving the bounds on the invisible decay rate of the Higgs.
 In principle, it can also be probed by studying the neutral current neutrino nucleus scattering experiments in which the neutrino
  beam is energetic enough to produce the sterile neutrinos.
   
   In the case of $\epsilon_{e \mu}$ and $\epsilon_{e \tau}$, we require a mixing between the sterile neutrinos with $\nu_e$. To avoid the strong bounds from neutrinoless double beta decay searches on the mixing of a Majorana sterile neutrino with $\nu_e$, we must adopt the lepton number conserving version of the model. In the case of $\epsilon_{\tau \mu}$,  it is also favorable to adopt the lepton number conserving version to avoid the constraint from searches for $B^+ \to \pi^+ \mu^- \mu^-$ by the LHCb \cite{Aaij:2014aba}.

   The couplings of the active neutrinos to $Z'$, $g_\nu$, can be as large as $10^{-3}$ constrained by the bounds on $K^+ \to \mu^+ Z' \nu$ \cite{Pouya1}. As shown in \cite{Pouya2} smaller values of $g_\nu$ can be tested by the near detector of DUNE. On the other hand, the couplings to quarks can also be as large as $10^{-3}$ constrained by $\pi^0 \to Z' \gamma$  for $m_{Z'}<m_{\pi^0}$ \cite{Gninenko:1998pm}. Within this part of the parameter space of our model, we can obtain values of $\epsilon_{\alpha \beta}^q$ saturating the present bounds on it.
   
   We have shown that despite LFV, at one loop level, there is no contribution to $l^-_\alpha \to l^-_\beta \gamma$ so no significant bound comes  even from the very stringent bound on  Br($\mu^-\to e^- \gamma$) on our model parameter space. However, at one loop level, there is a contribution to 
   $\mu^-$ to $e^-$ conversion on nuclei through virtual $Z'$ exchange which can be probed by COMET and mu2e experiments in future. Moreover, for $m_{Z'}<m_\mu$ at one loop level, we obtain $\mu^- \to e^- Z'$ whose effect may be probed by studying the energy spectrum of $e^-$ emitted in the muon decay.
   A signal for $\mu^--e^-$ conversion in the experiments without a signal for $\mu^- \to e^- \gamma$ in the future searches combined with  the traces of $Z'$ in the muon and
   meson decays along with observable $\epsilon_{e\mu}^f$ and a deviation of the Higgs invisible decay rate from the SM prediction can be considered smoking gun signatures for the present model.
 \acknowledgments
 This project has received funding from the European Union\'~\!s Horizon 2020 research and innovation programme under the Marie Sklodowska-Curie grant agreement No.~674896 and No.~690575. YF has received partial financial support from Saramadan under contract No.~ISEF/M/98223. YF would like also to thank the  ICTP staff and the INFN node of the INVISIBLES network in Padova. She is also grateful
 to the staff of Departament de Fisica Teorica of Valencia University, especially G. Barenboim, for their hospitality. 


\end{document}